\documentstyle[12pt]{article}
\title{Simple Formula for Nuclear Charge Radius \footnote{This work is partly
supported by the Polish Committee of Scientific Research under contract
No. 203119101.}}
\author{Bo\.zena Nerlo--Pomorska and Krzysztof Pomorski \\
Theoretical Physics Department, The Maria Curie--Sk\l odowska University,\\
Radziszewskiego 10, 20031 Lublin, Poland}

\begin{document}

\maketitle

\begin{abstract}

A new formula for the nuclear charge radius is proposed, dependent on the
mass number $A$ and neutron excess $N-Z$ in the nucleus. It is simple
and it reproduces all the experimentally available mean square radii and their
isotopic shifts of even--even nuclei much better than other frequently used
relations.

\end{abstract}

\newpage
\section{Introduction}

{}~~~~~The aim of the present investigation is to find a phenomenological
macroscopic formula for the nuclear radius.
A charge distribution inside a nucleus depends not only on it's mass number
$A$ and the ground state deformation caused by shell effects, but also on the
neutron excess $N - Z$. The equilibrium deformation of a nucleus can be
established by the minimalization of the potential energy obtained by one of
the microscopic methods. But to know the range ($R$) of nuclear charge in
any ($\vartheta, \varphi$) direction, it is necessary to describe properly
the radius ($R_{00}$) of the corresponding spherical nucleus which has been
deformed keeping its volume constant.

The simplest known liquid drop formula for the nuclear radius
\begin{equation}
R_{00}= r_0\cdot A^{1/3}
\end{equation}
with $r_0=1.2 fm$ doesn't lead to the experimentally known mean square radii
$\langle r^{2}\rangle$ (MSR) [1]
or their isotopic shifts [2]. A popular liquid droplet model [3]
gives also rather unsatisfactory results for MSR.

We are going to find the best dependence of $R_{00}$ on mass number ($A$)
and neutron excess ($N-Z$). A corresponding formula which
reproduces the experimentally measured isotopic shifts of MRS was already
proposed in refs. [4, 5]. That formula was obtained without the fit to the
absolute values of the model independent of MSR data [1]. Now we would like
to propose a new simple expression for the proton radius ($R_{00}$) which
reproduces not only the isotopic shifts but also the magnitude of MSR
for all even--even nuclei with proton numbers $Z \geq 8$.

\section{Description of the calculation}

{}~~~~~The MSR values of even--even nuclei are calculated macroscopically with
a uniform charge distribution
\begin{equation}
 \rho_{0} =
 \left\{
  \begin{array}{ccc}
   \frac{Ze}{V} & \mbox{for} & r \leq R \\
     0          & \mbox{for} & r > R
  \end{array}\right.
\end{equation}
where  $V$ is the volume of a nucleus conserved during deformation
\begin{equation}
 V = \frac{4}{3}\pi R^{3}_{00} = \int^{2\pi}_0 d\,\varphi
     \int^1_{-1} d\,\cos\vartheta
     \int^{R(\{\beta_\lambda\};\vartheta,\varphi)}_{0} r^{2}\,d\,r\,\,\,.
\end{equation}
$R$ is the radius dependent on deformation, calculated with the equilibrium
shapes $\{\beta_\lambda\}$ taken from ref. [6].

The MSR values are given by the integral
\begin{equation}
 \langle r^{2}\rangle = \rho_{0} \int^{2\pi}_{0} d\varphi \int^{1}_{-1}
  d \cos \vartheta \int^{R(\{\beta_{\lambda}\};\vartheta,\varphi)}_{0} r^{4} dr
\end{equation}
and they are evaluated for axially symmetric nuclei  with the radius
\begin{equation}
 R(\{\beta_{\lambda}\},\vartheta) = R_{0}(\{\beta_{\lambda}\})
\left[1 + \sum^{6}_{\lambda=1} \beta_{\lambda}P_{\lambda}(\cos\vartheta)\right]
  \,\,\, ,
\end{equation}
where $P_\lambda$ are the Legender polynomials and
\begin{equation}
 R_{0}(\{\beta_{\lambda}\}) = R_{00}\left\{\frac{1}{2} \int^{1}_{-1}
\left[1 + \sum^{6}_{\lambda=1} \beta_{\lambda}P_{\lambda}(\cos\vartheta)\right]
 d\cos\vartheta\right\}^{-1/3}  \,\,\,  .
\end{equation}
The last relation originates from the volume conservation condition when
a nucleus deforms and $R_{00}$ is the radius of the corresponding spherical
nucleus. For practical use the formula (4) for small
quadrupole deformations could be estimated approximatly by:
\begin{eqnarray}
 \langle r^{2}\rangle & = & \frac{Q_{00}} {Z} \left\{1 + \frac{8} {\sqrt{5\pi}}
  \beta^{2} + \cdots \right\} \nonumber \\
  & = & \frac{Q_{00}} {Z} \left\{1 + \frac{8 }{5\sqrt{5\pi}}
  \left(\frac{Q_{20}}{Q_{00}} \right)^{2} + \cdots \right\},
\end{eqnarray}
where $Q_{00} = \frac{3}{5} e\,Z\,R^{2}_{00}$ is the electric monopole moment
of the spherical nucleus and $Q_{20}$ is the electric quadrupole moment of the
deformed nucleus.

The charge radius $R_{00}$ in formula (6) is chosen in the following form
(see also [4-5])
\begin{equation}
 R_{00} = r_{0}A^{1/3} \left(1 + \frac{\kappa}{A} - \alpha \frac{N-Z}{A}\right)
\end{equation}
The set of $r_{0}$, $\kappa$ and $\alpha$ parameters are obtained by
the fit of $\langle r^{2}\rangle^{exp}$ [1] and its isotopic shifts
$\delta <r^2>^{A,A'}_{exp}$ [2] for even--even nuclei with the proton numbers
$Z \geq 8$. We have assumed that the sum of the mean square errors:
\begin{equation}
 \Sigma^{2}_{\langle r^{2}\rangle} = \kern-5mm
  \sum^{86}_{%
  \begin{array}{c}
  \scriptstyle i = 1\\ [-2dd]
  \scriptstyle Z \geq Z_{min}
  \end{array}}\kern-5mm
  \left(\langle r^{2}\rangle_{i} - \langle r^{2}\rangle^{exp}_{i}\right)
\end{equation}
should be minimal. Simultaneously we have kept the minimal sum of square errors
of MSR isotopic shifts for all experimentally measured sets of even--even
isotopes
\begin{equation}
 \delta\langle r^{2}\rangle^{A,A'} = \langle r^{2}\rangle^{A} -
 \langle r^{2}\rangle^{A'}
\end{equation}
i.e.
\begin{equation}
 \Sigma^{2}_{\delta\langle r^{2}\rangle} = \kern-5mm
  \sum^{220}_{%
  \begin{array}{c}
  \scriptstyle i = 1\\ [-2dd]
  \scriptstyle Z \geq Z_{min}
  \end{array}}\kern-5mm
  \left(\delta\langle r^{2}\rangle^{A,A'}_{i} - \delta\langle r^{2}
  \rangle^{A,A'}_{i,exp}\right) \,\,\,  .
\end{equation}
The results obtained with our simple formula for the charge radius (8)
will be compared in the next section with the liquid droplet estimates
of MSR [3].

\section{Results}

After an analysis of 22 isotopic shifts of MSR for even--even nuclei we
have proposed in Ref. [5] a formula for $R_{00}$ containing the isotopic term,
\begin{equation}
R_{00} = r_{0}A^{1/3}\left(1 - \alpha\frac{N-Z}{A}\right)
\end{equation}
with the parameters $\alpha = 0.2$ and $r_{0} = 1.25 fm$,
but without the $\sim 1/A$ term, i.e. $\kappa = 0$. This formula was good
enough to reproduce all the MSR isotopic shifts of even--even elements above
$Z \geq 38$ and the average magnitude of MSR for the gold isotopes.
Unfortunately it was rather impossible to get a reasonable agreement of MSR
values of lighter elements $8 \leq Z < 38$ without introducing an
additional $\kappa/A$ term. Fitting $R_{00}$ described by formula (6) to the
MSR [1] and $\delta\langle r^{2}\rangle^{A,A'}$ [2] data for all the
$Z \geq 8$ elements we have recorded the following set of parameters:
$$ r_{0} = 1.240 fm,\,\, \kappa = 1.646,\,\, \alpha = 0.191 $$ \,\, .
In every case the sum of square errors of MSR is more than six times lower than
in the liquid droplet model [3].
It is interesting, that the $\alpha$ parameter is almost stable around
$\alpha \sim 0.2$, independently on the number of fitted data and inclusion
of the $\sim 1/A$ term.
The values $r_{0}$ and $\alpha$ are almost the same for the case of $Z > 38$
elements without the $\kappa/A$ in $R_{00}$.

We have also tested the different powers of $A$ in the new term
$\kappa/A^{n}$ ($n = \frac{1}{3},\frac{2}{3},\cdots 2$). We have
achives the best agreement for $n=1$, but for $n = \frac{1}{3}$ and
$\frac{2}{3}$ the quality of the fitment was comparable with that for
$n = 1$.

The results of our investigation are illustrated in the figures 1-5.
In the Figs. 1-3 the results of the MSR values obtained in the three cases
of $R_{00}$ are illustrated.

In Fig. 1 the MSR values obtained with the traditional liquid drop formula (1)
are plotted. They are obviously to small
even for the spherical case (dashed line) in comparison with the
experimental data [1] marked by stars. The difference between the macroscopic
results including the effect of the deformations of nuclei taken from Ref. [6]
and the experimental data $\langle r^{2}\rangle_{exp}$ are represented
by the length of vertical lines shown under each experimentally known
element (''bars''). The differences are rather large, especially for the light
elements.

In Fig. 2 the MSR results were obtained using our previous formula [5], but
with the better set of parameters ($r_{0} = 1.256 fm$, $\alpha = 0.202$),
fitting, both $<r^2>$ and $\delta\langle r^{2}\rangle^{A,A'}$ values.
The fitment is much better than that in Fig. 1, both for heavier
and lighter elements, even in the spherical case. When including deformation,
the differences (bars on the figure) between macroscopic and experimental value
$\langle r^{2}\rangle_{th} - \langle r^{2}\rangle_{exp}$ almost disappear,
but the lighter elements are still not well reproduced.
It was necessary to introduce  an additional term in the formula for the
charge radius in order to increase the model estimates of MSR for the lightest
nuclei.

In Fig. 3 the MSR results for the formula (8) with the best set of parameters
fitted to the $\langle r^{2}\rangle_{exp}$ and
$\delta\langle r^{2}\rangle^{A,A'}_{exp}$ data are shown.
The agreement is good.  In spite of a few cases the differences between the
macroscopic and experimental MSR values have almost disappeared when the
deformations of the nuclides were taken into account [6].

One can compare these results with the liquid droplet calculation
according to ref. [3] illustrated in Fig. 4, where the theoretical results are
much smaller than the experimental data. Also the results for MSR obtained with
the newest version of the droplet model [6] are only a little better than those
of [3] in Fig. 4.
It is obvious that a new analysis of liquid droplet parameters in order
to reproduce MSR values is needed. We hope that our investigation will be
helpful.

The importance of introducing ($N - Z$) dependence in $R_{00}$ is illustrated
in Figs. 5-6, where the MSR isotopic shifts $\delta\langle r^{2}\rangle^{A,A'}
_{exp}$ of all the experimentally known even--even isotopes [2] are compared
with our estimates.

In Fig. 5 the differences between the macroscopic results calculated with
equilibrium deformations of [7]  and experimental data [1]
$$
\Delta\langle r^{2}\rangle^{A,A'}\equiv\delta\langle r^{2}\rangle^{A,A'}_{th}-
                                    \delta\langle r^{2}\rangle^{A,A'}_{exp}
$$
are plotted in the case of the traditional liquid drop formula (1).
The differences are large. They reach even $1\, fm^{2}$ for some nuclei.
When including the $R_{00}$ depending on isospin $\frac{N-Z}{A}$ (Fig. 6),
these differences decrease significantly.
One has to add that the same quality of the fit of
$\delta\langle r^{2}\rangle^{A,A'}$ as with formula (8) was already achieved
for the two parametric formula (12), but the magnitude of MSR values for light
nuclei are not well reproduced.

\section{Conclusions}

We can draw the following conclusions from our investigations:

\begin{itemize}

\item The traditional liquid drop formula, depending on mass number $A$ only,
      is not able to give a good mean square nuclear charge radius
      $\langle r^{2}\rangle$ or its isotopic shifts.

\item In order to reproduce the data of $\langle r^{2}\rangle$ and
      $\delta\langle r^{2}\rangle$ of the even--even nuclei one should use
      the three parametric formulas for the spherical nuclear charge radius
      $$
 R_{00} = 1.240\, A^{1/3}\left(1 + \frac{1.646}{A} + 0.191 \frac{N-Z}{A}\right)
      fm
      $$

\item For the nuclei with $Z \geq 38$ one can omit the $\sim 1/A$ term and
      use the formula
      $$
      R_{00} = 1.256\, A^{1/3} \left(1 - 0.202 \frac{N-Z}{A}\right) fm
      $$

\item The liquid droplet model gives systematically larger MSR
      estimates of the nuclear charge.

\end{itemize}
\vspace{0.5 cm}

The authors are very grateful to Prof. W. Swiatecki for his suggestions and
the helpful discussions. One of us (K.P.) would like also to thank the
Autonoma University in Madrid, where part of the investigation was completed,
for the good working conditions and the warm hospitality.

\vspace{1cm}

\section{References:}

\begin{enumerate}

\item H. de Vries, C. W. de Jager, C. de Vries, Atomic Data and Nucl. Data
      Tabl. {\bf 36} (1987) 495
\item E. W. Otten ''Treatise on heavy--ion science'', Vol. 8, Bromley D. A.
      (ed.), New York, Plenum Press (1989)
\item W. D. Myers, K.--H. Schmidt, Nucl. Phys. A410 (1983) 61

\item B. Nerlo--Pomorska, K.  Pomorski, Zeit. f\"ur Phys.{\bf A 344} (1993) 359

\item B. Nerlo--Pomorska, K. Pomorski, Proc.d of XXIIIth Masurian Lakes Summer
      School, Piaski, Poland, August 18-28 (1993), to be published in
      Acta Physica Polonica
\item P. M\"oller, J. R. Nix, W. D. Myers, W. J. Swiatecki -- submitted to
      Atomic Data and Nucl. Data Tabl. (1993)
\item B. Nerlo--Pomorska, B. Mach -- submitted to Atomic Data and Nucl. Data
      Tabl. (1993)

\end{enumerate}

\newpage

\section{Figures Captions}

\bigskip
\noindent

{\bf Fig. 1.} The mean square charge radii in $fm^{2}$ of even even nuclei
obtained without deformation (dashed line) with the traditional formula
liquid drop compared with experimental data [1] (crosses).
The differences between the macroscopic results evaluated
including equilibrium deformations [6] and the experimental data [1]
are illustrated by the bars around the ordinates line. The
macroscopic MSR values are to small.

\bigskip

{\bf Fig. 2.} The same as in Fig. 1 but with the isospin dependence in
the $R_{00}$ formula (12). The differences for light nuclei are still of
the order $\sim 1 fm^{2}$.

\bigskip

{\bf Fig. 3.} The same as in Fig. 1 but with our new formula for $R_{00}$
(8). The experimental data is almost ideally reproduced.

\bigskip

{\bf Fig. 4.} The same as in Fig. 1 for the liquid droplet macroscopic results
with parameters of Ref. [3]. The theoretical values are to small up to
$2 fm^{2}$.

\bigskip

{\bf Fig. 5.} The differences between the liquid drop and the
experimental isotopic shifts of the MSR of the charge in $fm^2$.
The discrepancy between these macroscopic results and the experimental
data exceed $1 fm^2$.
\bigskip

{\bf Fig. 6.} The same as in Fig. 5 for the three parametric $R_{00}$ formula
(7). The deviations between our theoretical estimates and experimental
data do not exceed $0.5 fm^{2}$.

\end{document}